\documentclass[aps,preprint,epsfig,rotate]{revtex4}
\usepackage{graphicx}
\usepackage{bm}
\usepackage{epsfig}

\input epsf
\begin{document}
\title{Highly accurate bound state calculations of the two-center molecular ions by using the universal variational expansion for three-body systems}

\author{Alexei M. Frolov}
 \email[E--mail address: ]{alex1975frol@gmail.com  and  afrolov@uwo.ca}

\affiliation{ITAMP, Harvard-Smithonian Center for Astrophysics, \\
         MS 14, 60 Garden Street, Cambridge MA 02138-1516, USA}  

\affiliation{Department of Applied Mathematics \\
 University of Western Ontario, London, Ontario N6H 5B7, Canada}

\date{\today}

\begin{abstract}

The universal variational expansion for the non-relativistic three-body systems is explicitly constructed. Three-body universal expansion can be 
used to perform highly accurate numerical computations of the bound state spectra in arbitrary three-body systems, including Coulomb three-body 
systems with arbitrary particle masses and electric charges. Our main interest is related to the adiabatic three-body systems which contain one 
bound electron and two heavy nuclei of hydrogen isotopes: the protium $p$, deuterium $d$ and tritium $t$. We also consider the analogous (model) 
hydrogen ion ${}^{\infty}$H$^{+}_2$ with the two infinitely heavy nuclei. 

\noindent 
PACS number(s): 36.10.-k and 36.10.Dr

\end{abstract}

\maketitle


\section{Introduction}

In this communication we investigate the bound state spectra in the non-relativistic three-body systems. Our main goal is to develop the universal variational 
expansion for arbitrary three-body systems, including different Coulomb three-body systems. Briefly, we want to develop the new fast and reliable approach 
which can be used for highly accurate computations of the total energies and various bound state properties of these three-body systems. First, let us consider
the Coulomb three-body systems with unit electrical charges \cite{FroB92}. Among such systems one finds three fundamental systems (ions): positronium ion 
Ps$^{-}$, hydrogen ion ${}^{\infty}$H$^{-}$ and hydrogen molecular ion ${}^{\infty}$H$^{+}_2$ which form a `natural' basis for the three-body systems with unit 
(electric) charges. The word `natural' used here means that: (1) these systems represents all three limiting cases of the Coulomb three-body systems with unit 
charges and different particle masses $m_1, m_2, m_3$, and (2) the energies and properties of these systems expressed in atomic units do not depend upon any 
additional `experimental' value of some particle mass (or masses). This fact was already mention in the first editions of some classical books on few-body 
systems (see, e.g., \cite{BS}, \cite{Eyr}). 

First highly accurate computations of the bound states in two-electron atoms and ions were performed in the middle of 1950's. Analogous calculations for the 
two-center adiabatic ions, e.g., H$^{+}_{2}$ ion started ten years later, while accurate and highly accurate variational calculations of the Coulomb three-body 
systems with three comparable particle masses, e.g., calculations of the Ps$^{-}$ ion and muonic molecular ions $pp\mu, pd\mu, dt\mu$, etc, began in the 
late 1970's. At that time for each Coulomb three-body system we applied that variational expansion which was specifically designed and/or modified for this 
system. In rare cases the same variational expansion could be used for highly accurate variational calculaltions of bound states in a number of similar systems, 
e.g., the both Hylleraas and exponential variational expansions were used to determine the bound state spectra in all six muonic molecular ions $pp\mu, pd\mu, 
pt\mu, dd\mu, dt\mu$ and $tt\mu$. Finally, in the middle of 1980's many different variational expansions have been developed and used for accurate numerical
computations of various three-body systems. For some of three-body systems these variational expansions produced very accurate results, but in applications to 
other similar systems were essentially useless. In addition to this, for a large number of Coulomb three-body systems no highly accurate variational expansion 
were known. On the other hand, at that time it was shown that a few simple modifications of the variational expansions developed earlier for three-body systems 
made them almost `universal', i.e. they can be applied for highly accurate computations of almost arbitrary three-body systems. The principal question here was 
to remove a few remaining trouble spots detected for such variational expansions. For instance, for the exponential variational expansion such a spot was 
associated with the molecular one-electron ions, i.e. with the $(a b e)^{+}$ ions, where $a$ and $b$ are the heavy nuclei of the hydrogen isotopes and $e$ is 
the electron. Finally, this was understood in \cite{Fro1987} and since then  the goal of numerous studies was to develop such a universal variational expansion 
for the three-body systems and, first of all, for the Coulomb three-body systems.      

In general, it would be a great advantage for the whole few-body physics to construct such an universal variational expansion, since this will allow one to 
perform highly accurate bound state computations for arbitrary three-body systems, including, three Coulomb systems/ions with unit charges (and arbitrary 
masses!) mentioned above. The explicit construction of the universal variational expansions is needed to investigate the bound state spectra, calculate the 
lowest-order relativistic and QED corrections to the non-relativistic energies and evaluate various properties of the non-relativistic three-body systems. In 
this study by the universal variational expansion we understand the variational expansion which allows one to perform highly accurate variational wave functions 
for an arbitrary bound state in any three-body (non-relativistic) system. It can be shown that the explicit construction of the universal variational expansion 
for arbitrary three-body systems can be reduced to the case of Coulomb three-body systems (ions) with unit charges $X^{+}Y^{+}Z^{-}$ \cite{FroB92}. In other 
words, an ability of some variational expansion for three-body systems to be `universal' substantially depends upon the masses of the three particles, and does 
not depend upon their electric charges. As follows from here the universal variational expansion must produce highly accurate energies and wave function for an 
arbitrary three-body system (ion) $X^{+} Y^{+} Z^{-}$, including three fundamental ions mentioned above.

Further investigation of this problem brought us to the conclusion that the `trouble spot' of many variational expansions developed for atomic three-body systems 
coincides with the so-called adiabatic systems, i.e. systems which contain the two heavy particles and one light particle. Adiabatic Coulomb three-body systems 
contain two very heavy atomic nuclei (or two centers) and one bound electron. In some studies these systems were called either the two-center systems, or molecular 
ions (see, e.g., \cite{FroS1995}), \cite{Fro2001A}. In 2002 we have we presented the universal variational expansion which provided highly accurate for the Coulomb 
three-body systems with unit charges \cite{Fro2002}. However, despite an obvious success of those studies \cite{Fro2001A}, \cite{Fro2002} for the H$^{+}_2$, 
HD$^{+}$, HT$^{+}$, D$^{+}_2$ DT$^{+}$ and T$^{+}_2$ ions some important questions of highly accurate computations of the adiabatic three-body systems have 
not been investigated in \cite{Fro2002}. In particular, the `pure adiabatic' ion ${}^{\infty}$H$^{+}_2$ was not considered in those works et al. However, this 
one-electron ion is one of the three fundamental three-body systems (ions) mentioned above. Also, in \cite{Fro2002} we have not constructed the highly accurate 
short-term wave functions with $N$ = 400 - 800 basis functions (complex exponents) for any of the H$^{+}_2$, HD$^{+}$, HT$^{+}$, D$^{+}_2$, DT$^{+}$ and T$^{+}_2$ 
ions. In addition to this, after a number of new experiments in high-energy physics the recommended numerical values of electron's mass and nuclear masses of the 
three hydrogen isotopes have been changed. The total energies of the bound states in the H$^{+}_2$, HD$^{+}$, HT$^{+}$, D$^{+}_2$ DT$^{+}$ and T$^{+}_2$ ions are
changed correspondingly, and we need to evaluate these changes to very high accuracy. Moreover, the overall accuracy of our results obtained in earlier studies 
\cite{Fro2002} was not high enough. Our current results obtained for these three-body ions are $\approx 2 \cdot 10^{4} - 5 \cdot 10^{5}$ times more accurate. This 
allows us to construct the bound state wave functions of the adiabatic $(a b e)^{+}$ ions which have significantly better overall quality. Taking into account all 
these issues we have decided to re-consider the adiabatic two-canter ions by using our universal expansion and answer these and other important questions. 

This paper has the following structure. The general Coulomb three-body problem is formulated in the next Section. In Section III we consider the `adiabatic 
divergence' of the three-body variational expansion(s) written in the relative (or Hylleraas) coordinates $r_{32}, r_{31}, r_{21}$. To analyze the source of this 
`adiabatic divergence' we present the explicit form of the Hamiltonian of the non-relativistic three-particle problem. Results of our numerical computations for 
the ions which contain two heavy positively charged nuclei of hydrogen isotopes the protium $p$, deuterium $d$ and tritium $t$ are analyzed in the fourth Section. 
The short term variational wave functions for these three-body adiabatic ions are considered in the fifth Section. The `pure adiabatic' ${}^{\infty}$H$^{+}_{2}$ 
ion is discussed in the sixth Section. Concluding remarks can be found in the last Section.

\section{Coulomb three-body problem}
 
Let us consider the non-relativistic system of three point particles which have electrical charges $q_1 e, q_2 e, q_3 e$ and masses $m_1, m_2, m_3$. To describe 
such a system we introduce three Cartesian vectors ${\bf r}_i$ (= $(x_{i}, y_{i}, z_{i})$ for $i$ = 1, 2, 3) which describe the positions of the particles. Three 
vectors ${\bf r}_{32}, {\bf r}_{21}$ and ${\bf r}_{13}$, where ${\bf r}_{ij} = {\bf r}_{i} - {\bf r}_{j} = -{\bf r}_{ji}$, form a triangle of particles. Each rib 
of this triangle has length $r_{ij} = \mid {\bf r}_i - {\bf r}_j \mid = r_{ji}$ which coincides with the scalar interparticle distance $r_{ij}$, or relative 
coordinate. The differential vector-operators $\nabla_{i}$ (Hamilton's $nabla$) are defined as follows (in Cartesian coordinates) $\nabla_{i} = \Bigl( 
\frac{\partial}{\partial x_{i}}, \frac{\partial}{\partial y_{i}}, \frac{\partial}{\partial z_{i}} \Bigr)$ and $i = 1, 2, 3$. The Hamiltonian of the three-body 
system of non-relativistic particles is written in the form
\begin{eqnarray}
  H = &-& \frac{\hbar^2}{2 m_e} \Bigl[ \frac{m_e}{m_1} \nabla^{2}_{1} + \frac{m_e}{m_2} \nabla^{2}_{2} + \frac{m_e}{m_3} \nabla^{2}_{3} \Bigr]  
  + \frac{q_1 q_2 e^{2}}{r_{21}} + V_{21}(r_{21}) \nonumber \\
 &+& \frac{q_1 q_3 e^{2}}{r_{31}} + V_{31}(r_{31}) + \frac{q_2 q_3 e^{2}}{r_{32}} + V_{32}(r_{32}) \label{Ham}
\end{eqnarray}
where $\hbar = \frac{h}{2 \pi}$ is the reduced Planck constant and $m_e$ is the electron's mass and $e$ is the absolute value of the electron's electric charge. The 
potentials $V_{ij}(r_{ij})$ describe the non-Coulomb parts of inter-particle interactions. If we assume that all non-Coulomb parts of the interaction potentials equal 
zero identically, then we deal with the Coulomb three-body problem. The Coulomb three-body problems are of great interest in a large number of applications including 
various problems from atomic, molecular and optical physics, astrophysics, solid state physics, plasma physics and other areas. In many of these problems it is very 
important to obtain the highly accurate eigenvalues (or total energies $E$) and corresponding highly accurate wave functions $\Psi$ for different bound states. The 
highly accurate bound state wave functions are later used in various applications, e.g., in calculations of the lowest-order relativistic and QED corrections, for 
numerical evaluation of bound state properties and also to determine the final state probabilities of different decays, processes and reactions in the Coulomb 
three-body systems. Below, we restrict ourselves to the consideration of the Coulomb three-body systems only. 

In general, the total energies $E$ and wave functions $\Psi$ are determined as the solutions of the corresponding Schr\"{o}dinger equation $H \Psi = E \Psi$ for the 
bound states (i.e. $E < 0$), where $H$ is the Hamiltonian of the three-body system, Eq.(\ref{Ham}). As follows from Eq.(\ref{Ham}) the Hamiltonian of any Coulomb 
three-body systems has a simple form in the system of units where $\hbar = 1, m_e = 1$ and $e^2 = 1$. These units are the atomic units, and they are employed in 
atomic physics, since the middle of 1920's. In our study the atomic units will be used extensively. In these units the Hamiltonian, Eq.(\ref{Ham}), takes the form
\begin{eqnarray}
 H = - \frac{1}{2 m_1} \nabla^{2}_{1} - \frac{1}{2 m_2} \nabla^{2}_{2} - \frac{1}{2 m_3} \nabla^{2}_{3} + \frac{q_1 q_2}{r_{21}} + \frac{q_1 q_3}{r_{31}} + 
 \frac{q_2 q_3}{r_{32}} \label{Ham1}
\end{eqnarray}
where the first three terms are the single-particle kinetic energies, while the last three terms are the potential (or Coulomb) energies of the interparticle 
interaction(s). As is well known the Schr\"{o}dinger equation for the bound state spectrum can be derived from a variational principle which is applied to the energy 
functional $E = \langle \Psi \mid H \mid \Psi \rangle$, where $H$ is the Hamiltonian from Eq.(\ref{Ham1}) and the varied wave function $\Psi$ has a unit norm. In 
actual applications the `exact' wave function $\Psi$ is approximated by some variational expansion(s). In general, each of the highly accurate wave functions includes 
a large number $N$ of basis functions \cite{Eps} which explicitly depend upon spatial and spin coordinates of all three particles. There are nine spatial coordinates 
for an arbitrary three-particle system, but three of them describe the equal spatial translations of the particles, while three other coordinates represent the 
rotation(s) of the triangle of particles in our three-dimensional space. During such rotations the triangle of particles is considered as one solid formation. The 
six coordinates (which represent spatial translations and rotations) can be separated and excluded from the following analysis, e.g., by employing the three special 
coordinates which are translationally and rotationally invariant, i.e. they do not change during any translation and/or rotation in three-dimensional space. Three 
remaining coordinates are the radial (or internal) scalar coordinates of the problem. The correct choice of three internal coordinates is of paramount importance, 
since these coordinates determine the actual convergence rate of the variational expansion.

For the two-electron atoms and ions it was shown by Hylleraas \cite{Hyl1}, \cite{Hyl2} that the the relative coordinates $r_{32}, r_{31}$ and $r_{21}$, where $r_{ij} 
= \mid {\bf r}_{i} - {\bf r}_{j} \mid = r_{ji}$ provide the correct choice for the internal coordinates. Indeed, these three relative coordinates are rotationally and 
translationally invariant. This substantially simplifies all numerical computations of different bound states in atomic two-electron systems. Later, the approach 
developed by Hylleraas and other similar approaches have been generalized for three-body systems with three arbitrary particle masses, which, however, were always 
comparable to each other, e.g., for the Ps$^{-}$ ion \cite{Hyl2}, $pp\mu$ muonic molecular ion \cite{DK} 
and other similar systems (see, e.g., \cite{FroE1984}, \cite{BD1984}). Based on the results of \cite{DK} we have developed a modification of the three-body exponential 
expansion which was very useful for the investigation of the bound state spectra of various muonic molecular ions, including different `rotationally' and `vibritionally' 
excited states in such systems. This variational expansion is written in one of the two following forms (see, e.g., \cite{Fro2002}, \cite{FrWa2011} and references therein)
\begin{eqnarray}
 \Psi_{LM} = \frac{1}{2} (1 + \kappa \hat{P}_{21}) \sum_{i=1}^{N} \sum_{\ell_{1}} C_{i} {\cal Y}_{LM}^{\ell_{1},\ell_{2}} ({\bf r}_{31},
 {\bf r}_{32}) \exp(-\alpha_{i} r_{32} - \beta_{i} r_{31} - \gamma_{i} r_{21})  \label{equ5} 
\end{eqnarray}
which is called the three-body exponential variational expansion in the relative coordinates $r_{32}, r_{31}, r_{21}$, or 
\begin{eqnarray}
 \Psi_{LM} = \frac{1}{2} (1 + \kappa \hat{P}_{21}) \sum_{i=1}^{N} \sum_{\ell_{1}} C_{i} {\cal Y}_{LM}^{\ell_{1},\ell_{2}} ({\bf r}_{31},
 {\bf r}_{32}) \exp(-\alpha_{i} u_1 - \beta_{i} u_2 - \gamma_{i} u_3)  \label{equ6} 
\end{eqnarray}
which is called the three-body exponential variational expansion in the perimetric coordinates $u_{1}, u_{2}, u_{3}$. The perimetric coordinates have been introduced 
in physics of three-body systems by C.L. Pekeris in \cite{Pek1}, \cite{Pek2}. These three coordinates are simply (linearly) related to the relative coordinates
\begin{eqnarray}
  & & u_1 = \frac12 ( r_{21} + r_{31} - r_{32}) \; \; \; , \; \; \; r_{32} = u_2 + u_3 \nonumber \\
  & & u_2 = \frac12 ( r_{21} + r_{32} - r_{31}) \; \; \; , \; \; \; r_{31} = u_1 + u_3 \; \; \; \label{coord} \\
  & & u_3 = \frac12 ( r_{31} + r_{32} - r_{21}) \; \; \; , \; \; \; r_{21} = u_1 + u_2 \nonumber
\end{eqnarray}
where $r_{ij} = r_{ji}$. In contrast with the relative coordinates $r_{32}, r_{31}, r_{21}$ the three perimetric coordinates $u_1, u_2, u_3$ are truly independent of each 
other and each of them varies between 0 and $+\infty$. This drastically simplifies analytical and numerical computations of all three-body integrals which are needed for 
solution of the corresponding eigenvalue problem and for evaluation of a large number of bound state properties in an arbitrary three-body system.

In Eqs.(\ref{equ5}) - (\ref{equ6}) the coefficients $C_{i}$ (everywhere below we assume that $i = 1, 2, \ldots, N$) are the linear variational parameters, while the 
$\alpha_i, \beta_i, \gamma_i$ are the non-linear (variational) parameters and $L$ is the angular momentum of the three-body system. The functions 
${\cal Y}_{LM}^{\ell_{1},\ell_{2}} ({\bf r}_{31}, {\bf r}_{32})$ in Eqs.(\ref{equ5}) - (\ref{equ6}) are the bipolar harmonics \cite{Varsh}, which depend upon the two 
vectors ${\bf r}_{31} = r_{31} \cdot {\bf n}_{31}$ and ${\bf r}_{32} = r_{32} \cdot {\bf n}_{32}$, where the ${\bf n}_{3i}$ vectors are the corresponding unit vectors ($i$ 
= 1, 2). The explicit form of the bipolar harmonics is \cite{Varsh}
\begin{equation}
 {\cal Y}_{LM}^{\ell_{1},\ell_{2}} ({\bf x}, {\bf y}) = x^{\ell_{1}} y^{\ell_{2}} \sum_{\ell_{1}, \ell_{2}} C^{LM}_{\ell_{1} m_{1};\ell_{2} m_{2}} 
 Y_{\ell_{1} m_{1}} ({\bf n}_{x}) Y_{\ell_{2} m_{2}} ({\bf n}_{y}) \label{e7}
\end{equation}
where $C^{LM}_{\ell_{1} m_{1};\ell_{2} m_{2}}$ are the Clebsch-Gordan coefficients (see, e.g., \cite{Varsh} and \cite{Rose}). As follows from Eq.(\ref{e7}) each bipolar 
harmonic is the $M-$component of the irreducible tensor of rank $L$. In actual calculations it is possible to use only those bipolar harmonics for which $\ell_{1} + 
\ell_{2} = L$ \cite{FroE1984}. This menas that each basis function in Eq.(\ref{equ6}) is an eigenfunction of the $L^2$ and $L_z$ operators with eigenvalues $L(L+1)$ and 
$M$. This means that $\hat{L}^2 \Psi_{LM} = L (L + 1) \Psi_{LM}$, while $M$ is the eigenvalue of the $\hat{L}_z$ operator, i.e. $\hat{L}_z \Psi_{LM} = M \Psi_{LM}$. The 
operator $\hat{P}_{21}$ in Eq.(\ref{equ6}) is the permutation of the two identical particles in symmetric three-body systems. For such systems in Eqs.(\ref{equ5}) - 
(\ref{equ6}) one finds $\kappa = \pm 1$, otherwise $\kappa = 0$. 

Note that the two expansions, Eqs.(\ref{equ5}) - (\ref{equ6}), are the partial cases of the more general exponential variational expansion written in the relative/perimetric 
coordinates \cite{FrWa2011}. In particular, the expansions, Eqs.(\ref{equ5}) - (\ref{equ6}), do not include additional factors which can be used to accelerate the overall 
convergence rate for some special three-body systems (for more details, see, \cite{Fro2002} and \cite{FrWa2011}). In this study we restrict ourselves to the bound (three-body) 
states with $L = 0$. In this case the explicit form of the exponential variational expansion, Eq.(\ref{equ6}), is simplified to the form 
\begin{eqnarray}
 \Psi(u_1, u_2, u_3) = \frac{1}{2} (1 + \kappa \hat{P}_{21}) \sum_{i=1}^{N} C_{i} \exp(-\alpha_{i} u_1 - \beta_{i} u_2 - \gamma_{i} u_3)  \label{equ75} 
\end{eqnarray}
where the non-linear parameters $\alpha_{i}, \beta_{i}$ and $\gamma_{i}$ ($i$ = 1, 2, $\ldots, N$) are the real/complex numbers which have non-zero imaginary parts. This 
variational expansion, Eq.(\ref{equ75}), is applied below. This variational expansion, Eq.(\ref{equ75}), is often written in a slightly different (but equivalent!) form 
\begin{eqnarray}
 \Psi(u_1, u_2, u_3) = \frac{1}{2} (1 + \kappa \hat{P}_{21}) \sum_{i=1}^{N} C_{i} \exp(-\alpha_{i} u_1 - \beta_{i} u_2 - \gamma_{i} u_3 - \imath \delta_{i} u_1 - \imath 
 e_{i} u_2 - \imath f_{i} u_3)  \label{equ755} 
\end{eqnarray}
where all $6 N-$non-linear parameters $\alpha_{i}, \beta_{i}, \ldots, f_i$ ($i = 1, \ldots, N$) are now considered to be real. In actual applications, the three last non-linear
parameters (i.e. the $\delta_i, e_i, f_i$ parameters) in each of the basis function in Eq.(\ref{equ755}) can be chosen as arbitrary real numbers (positive, or negative), while 
the first three non-linear parameters $\alpha_{i}, \beta_{i}, \gamma_{i}$ must always be positive (real) numbers. The radial set of exponential basis functions must be complete. 
From here one finds a set of three additional conditions for the $\alpha_{i}, \beta_{i}, \gamma_{i}$ parameters. Indeed, the three series of inverse powers of these parameters 
must be divergent, i.e. the three following sums (or series): $S_1 = \sum_{i=1} \frac{1}{\alpha_{i}}, S_2 = \sum_{i=1} \frac{1}{\beta_{i}}$ and $S_3 = \sum_{i=1} 
\frac{1}{\gamma_{i}}$ must diverge, if $i \rightarrow \infty$.

\section{Adiabatic divergence}

In the middle of 1980's we have developed an effective technique which allowed us to perform highly accurate bound state computations of arbitrary, in principle, Coulomb 
three-body system by using the exponential variational expansion, Eq.(\ref{equ75}), with the real non-linear parameters only. Let $E(N)$ be the total energy of the three-body 
system obtained with the use of $N$ basis functions, Eq.(\ref{equ75}). For a large number Coulomb three-body systems the actual convergence of the total energies $E(N)$ be 
represented by the following approximate formula
\begin{eqnarray}
     E(N) = E(\infty) + \frac{A}{N^{\nu}} \label{asymp}
\end{eqnarray}
where $N$ is the total number of basis functions used in Eq.(\ref{equ75}), while $E(\infty), A (\ge 0)$ and $\nu (\ge 1)$ are the varied `experimental parameters' which are 
determined from the results of numerical calculations. The numerical value of the parameter $\nu$ in Eq.(\ref{asymp}) is crucial for the whole method, since it determines the 
overall convergence rate of the variational expansion. Originally, the formula, Eq.(\ref{asymp}), and other similar formulas were used to produce a few additional correct 
decimal digits in the final result/energy, since the $E(\infty)$ energy approximates the total energy more accurately than any of the partial energies $E(N)$. This is the main 
reason why the formula, Eq.(\ref{asymp}), has been called the `asymptotic' formula for the total energy $E$. Analogous formulas can be derived for other bound state properties.

First applications of the formula, Eq.(\ref{asymp}), (with the real non-linear parameters) to the three-body muonic ions $pp\mu, dd\mu$ and $tt\mu$ computed with the use of 
Eq.(\ref{equ75}) indicated clearly that the numerical value of the parameter $\nu$ rapidly decreases when the mass of the heavy particle in the $ a a \mu$ ion (where $a = (p, d, 
t$)) increases. Indeed, the typical values of $\nu$  for the $pp\mu$ ion were around 10.5 - 11.5, while for the $dd\mu$ ion $\nu \approx$ 8.0 - 8.75 and for the $tt\mu$ ion 
$\nu \approx$ 5.3 - 6.1. In other words, the factor $\nu$ in Eq.(\ref{asymp}) is substantially mass dependent, i.e. it depends upon the ratio of particle masses. Formally, the 
overall accuracy of the variational expansion, Eq.(\ref{equ75}), with the real non-linear parameters was sufficient to obtain highly accurate results even for the $t t \mu$ 
ion. However, the rapidly decreasing value of $\nu$ in Eq.(\ref{asymp}) indicated that our variational expansion, Eq.(\ref{equ75}), with the real non-linear parameters is 
useless for highly accurate computations of the adiabatic one-electron ions such as $(p p e)^{+}, (p d e)^{+} , (p t e)^{+}, \ldots, (t t e)^{+}$, etc. Accurate computations of 
the bound states in the truly adiabatic ${}^{\infty}$H$^{+}_{2}$ ion with the use of Eq.(\ref{equ75}) seem to be absolutely unrealistic, since for this ion the parameter 
$\gamma$ in Eq.(\ref{asymp}) was less than unity. In reality, for the ground state of the ${}^{\infty}$H$^{+}_{2}$ ion it was hard to obtain even two correct decimal digits 
(see, e.g., \cite{Fro1987}), if all non-linear parameters in Eq.(\ref{equ75}) were real. This phenomenon was called \cite{Fro1987} the `adiabatic divergence' of the variational 
expansion, Eq.(\ref{equ75}). The same conclusion can be made for other variational expansions in the relative coordinates $r_{32}, r_{31}, r_{21}$ originally developed for 
highly accurate calculations of the two-electron atomic systems.        

Sources of the adiabatic divergence in three-body systems can be understood by considering the Hamiltonians of such systems. Formally, our analysis can be generalized to the 
adiabatic few- and many-body systems and for arbitrary interaction potentials between particles. However, such an analysis of arbitrary few-body systems just complicates the 
notations, and below we restrict ourselves to the consideration of the Coulomb three-body systems with the unit electric charges. Below, such systems are designated as 
$(a b e)$, or $(a b e)^{+}$, where $e$ is the electron, while $a$ and $b$ are the two heavy nuclei of hydrogen isotopes. Furthermore, for simplicity, we shall discuss the bound 
states in these systems with $L = 0$. The non-relativistic Hamiltonian $H$ of the $(a b e)^{+}$ system in the relative coordinates takes the form (in atomic units)
\begin{eqnarray}
 H = - \Bigl(\frac{1}{2 m_3} + \frac{1}{2 m_2}\Bigr) \Bigl[ \frac{\partial^2}{\partial r^{2}_{32}} + \frac{2}{r_{32}} \frac{\partial}{\partial r_{32}} \Bigr]
     - \frac{1}{2 m_3} \Bigl[\frac{r^{2}_{32} + r^{2}_{31} - r^{2}_{21}}{r_{32} r_{31}}\Bigr] \frac{\partial^2}{\partial r_{32} \partial r_{31}} \nonumber \\
     - \Bigl(\frac{1}{2 m_3} + \frac{1}{2 m_1}\Bigr) \Bigl[ \frac{\partial^2}{\partial r^{2}_{31}} + \frac{2}{r_{31}} \frac{\partial}{\partial r_{31}} \Bigr]
     - \frac{1}{2 m_2} \Bigl[\frac{r^{2}_{32} + r^{2}_{21} - r^{2}_{31}}{r_{32} r_{21}}\Bigr] \frac{\partial^2}{\partial r_{32} \partial r_{21}} \nonumber \\
     - \Bigl(\frac{1}{2 m_1} + \frac{1}{2 m_2}\Bigr) \Bigl[ \frac{\partial^2}{\partial r^{2}_{21}} + \frac{2}{r_{21}} \frac{\partial}{\partial r_{21}} \Bigr]
     - \frac{1}{2 m_1} \Bigl[\frac{r^{2}_{21} + r^{2}_{32} - r^{2}_{31}}{r_{21} r_{32}}\Bigr] \frac{\partial^2}{\partial r_{31} \partial r_{21}} \nonumber \\
     - \frac{1}{r_{32}} - \frac{1}{r_{32}} + \frac{1}{r_{32}} \; \; \; \label{Hamil}
\end{eqnarray}
where the particles 1 and 2 are the two heavy nuclei $p, d, t$ (nuclei of the hydrogen isotopes), while particle 3 is the electron ($e^{-}$). As follows from Eq.(\ref{Hamil})
in the limit when $m_1 \rightarrow \infty$ and $m_2 \rightarrow \infty$ (or $\min (m_1,m_2) \rightarrow \infty$) all terms which include derivatives in respect to the $r_{21}$
variable, i.e. $\frac{\partial}{\partial r_{21}}$ and/or $\frac{\partial^2}{\partial r^{2}_{21}}$ vanish from the Hamiltonian, Eq.(\ref{Hamil}). This means that in the adiabatic 
limit, i.e. when $\min (m_1,m_2) \rightarrow \infty$, the internuclear variable $r_{21} = R$ becomes an additional parameter of the three-body problem, i.e. it does 
not change during any electron's motion and cannot be considered as an actual coordinate in this problem. 

The fact that the variable $r_{21}(= R)$ is a constant during the electron's motion for the adiabatic (or two-center) systems means that we have to re-formulate the original 
variational three-body problem for the bound states to the following form. We need to find the unit-norm wave function $\Psi$ which provides a minimum to the following energy 
functional
\begin{eqnarray}
   E = \langle \Psi \mid H \mid \Psi \rangle \; \; \; \label{varia}
\end{eqnarray}
and obey the following additional conditions (or constraints) $\langle \Psi \mid r^{n}_{21} \mid \Psi \rangle = R^{n}$ for $n = 1, 2, \ldots$. Those wave functions which provide the
absolute minimum of the energy functional, Eq.(\ref{varia}), but do not obey these additional conditions have no actual physical meaining for the ${}^{\infty}$H$^{+}_2$ ion and close 
systems. Direct solution of the variational problem, Eq.(\ref{varia}), with an infinite number of additional constraints is a very difficult task. However, we can simplify this 
problem by assuming that the `exact' wave function $\Psi(r_{32}, r_{31}, r_{21})$ of the truly adiabatic system is represented in the form 
\begin{eqnarray}
   \Psi(r_{32}, r_{31}, r_{21}) = \delta(r_{21} - R) \phi(r_{32}, r_{31}) \; \; \; \label{phi}
\end{eqnarray}
where $\phi(r_{32}, r_{31})$ is the electron-nuclear part (or regular part) of the total three-particle wave function $\Psi$. Note that the explicit form of the `exact wave function, 
Eq.(\ref{varia}),  is written for the truly adiabatic two-center systems, e.g., for the ${}^{\infty}$H$^{+}_{2}$ ion. For actual three-body systems with the two heavy centers and one 
bound electron the exact delta-function in Eq.(\ref{phi}) must be replaced by some $\delta-$like spatial distribution, e.g., by the following one-parametric function
\cite{Sokolov}
\begin{eqnarray}
   \delta(r_{21} - R) \approx \frac{A}{\pi [A^2 + (r_{21} - R)^2]} \; \; \; \label{delta}
\end{eqnarray}
where $A$ is a numerical parameter. In the limit $A \rightarrow 0$ this function approaches to the actual two-center delta-function. 

Now, we can formulate the complete variational bound state problem for an arbitrary two-center three-body system in the form: {\it find the best approximation to the actual wave 
function which provides a minimum to the energy functional}, Eq.(\ref{varia}), {\it by using the trial functions which are represented in the form}
\begin{eqnarray}
   \Psi(r_{32}, r_{31}, r_{21}) = \frac{A}{\pi [A^2 + (r_{21} - R)^2]} \phi(r_{32}, r_{31}) \approx \delta(r_{21} - R) \phi(r_{32}, r_{31}) \; \; \; \label{phi1}
\end{eqnarray}
where $A$ is a small (real) number and $\phi(r_{32}, r_{31})$ is the regular electron-nuclear part of the three-body wave function. In other words, by using our `universal' 
variational expansion for non-relativistic three-body systems we need to find the best variational approximation of the wave function represented in the form, Eq.(\ref{phi1}). In 
the case of Eq.(\ref{equ75}) we can write this condition in the form 
\begin{eqnarray}
   \frac{A}{\pi [A^2 + (r_{21} - R)^2]} \phi(r_{32}, r_{31}) \approx \sum_{i=1}^{N} C_{i} \exp(-\alpha_{i} u_1 - \beta_{i} u_2 - \gamma_{i} u_3) \; \; \; \label{phi2}
\end{eqnarray}
where $A \rightarrow 0$ and all $\alpha_{i}, \beta_{i}, \gamma_{i}$ (for $i = 1, \ldots, N$) are real. It can be shown that the highly accurate numerical approximation of the regular 
$\phi(r_{32}, r_{31})$ function by the exponential variational expansion with the real non-linear parameters $\alpha_{i}, \beta_{i}$ and $\gamma_{i}$ in Eq.(\ref{phi2}) is a relatively 
simple problem. However, the same variational expansion, Eq.(\ref{phi2}), converges very slow for the $\delta-$like part of the total wave function. In fact, the overall convergence 
rate of the expansion which appears in the right-hand side of Eq.(\ref{phi2}) rapidly decreases when the parameter $A$ in Eq.(\ref{phi2}) approaches zero. This follows from the known 
fact that the Laplace transform of the delta-function(s) converges very slow, if all non-linear parameters in Eq.(\ref{phi2}) are real. The way to correct this situation is to apply 
the Fourier integral transform (or Laplace transform with the complex non-linear parameters) of the three-body wave function. In this case the convergence rate of the 
$\delta(r_{21} - R)$ factor represented by the total wave function, Eq.(\ref{delta}), increases drastically and we can produce the highly accurate variational wave functions even for 
pure adiabatic systems. Finally, the original variational problem, Eq.(\ref{varia}), with an infinite number of additional constraints is reduced to the minimization of the same 
energy functional with the use of the variational functions $\Psi$ which are represented in the two following equivalent forms: 
\begin{eqnarray}
  & & \Psi(r_{32}, r_{31}, r_{21}) = \frac{A}{\pi [A^2 + (r_{21} - R)^2]} \phi(r_{32}, r_{31})  \; \; \; \label{equa755} \\
 &\approx& \sum_{i=1}^{N} C_{i} \exp(-\alpha_{i} u_1 - \beta_{i} u_2 - \gamma_{i} u_3 - \imath \delta_{i} u_1 - \imath e_{i} u_2 - \imath f_{i} u_3) \nonumber 
\end{eqnarray}
where the last term is the variational series of complex exponents, Eq.(\ref{equ755}), while the parameter $A$ in Eq.(\ref{equa755}) is small and $A \rightarrow 0$. Generalization 
of this principle for the excited bound states in the adiabatic three-body systems is straightforward.  

The discussion above indicates that we have re-discovered the Born-Oppenheimer approximation for the two-center molecular ions \cite{BO}, \cite{Beth}. However, now the 
Born-Oppenheimer approach takes the form of a useful guide which can be used to construct the best variational wave functions for the adiabatic molecular ions, e.g., for three-body 
molecular ions such as $(p p e)^{+}, (p d e)^{+}, (p t e)^{+}, (d d e)^{+}, (d t e)^{+}$ and $(t t e)^{+}$. In our previous studies \cite{Fro2002} these systems have been designated 
as the H$^{+}_2$, HD$^{+}$, HT$^{+}$, D$^{+}_2$, DT$^{+}$ and T$^{+}_2$ ions, respectively. The method developed in this study allows one to produce very accurate values of the total 
and binding energies, variational wave functions and expectation values of different bound state properties of arbitrary three-body systems. Another interesting system for which the 
exponential variational expansion, Eq.(\ref{phi2}), can be applied is the truly adiabatic ${}^{\infty}$H$^{+}_{2}$ ion.  

\section{Numerical calculations}

Let us consider the ground bound states (or $1 s \sigma-$states) in one-electron $(p p e)^{+}, (p d e)^{+}, (p t e)^{+}, (d d e)^{+}, (d t e)^{+}$ and $(t t e)^{+}$ ions. It is 
clear that the computed value of the total energy depends upon the electron-nucleus mass ratio(s) used in calculations. In our calculations performed in this study we have used the 
two different sets of particle masses which can be found in modern literature. The first set of particle masses includes masses which have been obtained in recent high-energy 
experiments, i.e.
\begin{eqnarray}
   M_p = 938.272046 \; \; \; , \; \; \; M_d = 1875.612906 \; \; \; , \; \; \; M_t = 2808.920906 \label{eq001} 
\end{eqnarray}
where all masses are expressed in the mass-units $MeV/c^2$ accepted in the high-energy physics. The electron's mass $m_e$ in these units equals 0.510998910 $MeV/c^2$. From these 
mass values one can find the actual masses which have been used in our calculations. These masses form the set of `new' particle masses which are used below for a number of systems. 
An alternative set of particle masses which have extensively been used in earlier studies (see, e.g., \cite{Fro2002}) includes the following masses:
\begin{eqnarray}
   M_p = 1836.152701 m_e \; \; \; , \; \; \; M_d = 3670.483014 m_e \; \; \; , \; \; \; M_t = 5496.92158  m_e \label{eq01} 
\end{eqnarray}
where $m_e$ is the electron's mass. This set of particle masses is called below the set of `old' particle masses, or `old set of masses', for short. 

Our computational results of the total energies of the ground $1 s \sigma-$states in the $(p p e)^{+}, (p d e)^{+}, (p t e)^{+}, (d d e)^{+}, (d t e)^{+}$ and $(t t e)^{+}$ ions 
(or H$^{+}_2$, HD$^{+}$, HT$^{+}$, D$^{+}_2$ DT$^{+}$ and T$^{+}_2$) ions can be found in Tables I (old masses) and II (new masses). All results/energies in these Tables are given 
in atomic units. Results from these two Tables can be used to evaluate the corresponding convergence rate for each of the one-electron, two-center $(a b e)^{+}$ ions. As mentioned 
above such a convergence rate is mainly determined by the convergence rate of the internuclear (or two-center) delta-function expressed in terms of our variational expansion 
Eq.(\ref{equ755}). Again, we have to note that the exact delta-function $\delta(r_{21} - R)$ can be found in the truly adiabatic two-center molecular ions with the infinitely 
heavy nuclei, e.g., in the ${}^{\infty}$H$^{+}_{2}$ ion considered below. In actual one-electron molecular ions, e.g., in the $(p p e)^{+}, (p d e)^{+}, (p t e)^{+}, (d d e)^{+}, 
(d t e)^{+}$ and $(t t e)^{+}$ ions, we deal with the distributed (or delocalized) nucleus-nucleus delta-function, i.e. the delta-function of the distance between the two heavy 
nuclei. It is clear that for the $(a b e)^{+}$ ions with two light nuclei, e.g. for the $(p p e)^{+}$ and $(p d e)^{+}$ ions, the convergence rate of our variational expansion, 
Eq.(\ref{equ755}), is very high, while for analogous systems with heavy nuclei, e.g., for the $(d t e)^{+}$ and $(t t e)^{+}$ ions, this convergence rate slows down. Nevertheless, 
even for the $(d t e)^{+}$ and $(t t e)^{+}$ ions our variational expansion, Eq.(\ref{equ755}), produces highly accurate energies and wave functions. 

\section{Short-term cluster wave functions}

Another interesting question which we wanted to discuss in this study is the explicit construction of the short-term cluster wave functions for the adiabatic two-center molecular 
ions and other similar systems. In general, such short-term wave functions contain relatively small number of basis functions and each of these basis functions includes the same 
number of carefully optimized non-linear parameters. For the systems considered in this study this number equals six. The overall accuracy of such short-term wave functions is 
usually high, or very high, since even complete and accurate optimization of all non-linear parameters does not take long. Such functions are very convenient for fast and accurate 
evaluations of many bound state properties of these systems. Since \cite{Fro1998} the short-term cluster wave functions are extensively used to construct extremely accurate 
variational wave functions which contain significantly larger number of terms (or basis functions). 

The main idea of our two-stage optimization developed in \cite{Fro1998} and \cite{Fro2001} is simple and transparent: the unknown wave functions with the large number of basis 
functions $N$ is represented in the form $\Psi(N) = \Psi(N_0) + \Psi(N - N_{0})$, where $\Psi(N_0)$ is the short-term (cluster) wave function which contains a few hundreds 
carefully optimized non-linear parameters, while $\Psi(N - N_{0})$ is the wave function which contains a large number of basis functions, but explicitly depends upon fewer 
non-linear parameters (usually, a few dozens non-linear parameters). Very careful optimization of all non-linear parameters in the both $\Psi(N_0)$ and $\Psi(N - N_{0})$ functions 
allows one to obtain extremely accurate wave function(s) of arbitrary, in principle, bound state in any three-body system. 

In our earlier studies of the two-center adiabatic one-electron systems (ions) $(a b e)^{+}$we did not construct the short-term cluster wave functions \cite{Fro2002}. Here we want 
to present the highly accurate short-term wave functions for each two-center ion mentioned in this study. In other words, the short-term wave functions have been constructed for 
each of the $(p p e)^{+}, (p d e)^{+}, (p t e)^{+}, (d d e)^{+}, (d t e)^{+}, (t t e)^{+}$ ions and for the truly adiabatic hydrogen molecular ion ${}^{\infty}$H$^{+}_2$ with the 
two infinitely heavy nuclei. Results of our calculations of the total energies (in atomic units) for each of these ions with the short-term wave functions can be found in Table 
III. For each two-center ion with the finite nuclear masses in Table III we present the results obtained with the short-term wave functions with $N_0 = 400$ basis functions, 
Eq.(\ref{equ755}), while for the truly adiabatic ${}^{\infty}$H$^{+}_2$ ion we also want to show the variational energies obtained with the short-term functions which include 
$N_0$ = 200, 400, 600, 700 and 800 basis functions, Eq.(\ref{equ755}). The short-term wave functions with $N$ = 400, obtained in these calculations, have been used in the 
large-scale calculations of the three-body adiabatic ions with $N \ge 2000$ basis wave functions (see Tables I, II and IV). In general, for each three-body adiabatic ion the 
results obtained with these short-term wave functions are very close to the final energies obtained in computations of these ions with the larger numbers of basis functions.      

\section{Hydrogen ion with the infinitely heavy nuclei} 

In this Section we consider the one-electron H$^{+}_{2}$ ion with the infinitely heavy nuclei, i.e the ${}^{\infty}$H$^{+}_{2}$ ion. This model ion is the truly adiabatic 
two-center system, since the positions of the two atomic nuclei in this ion are strictly fixed and they do not change during any possible electron motion. As mentioned above 
highly accurate computations of the truly adiabatic two-center systems with the use of the three-body variational expansions originally developed for atomic one-center systems 
are always difficult to perform. The source of troubles for such expansions is obvious, since the ${}^{\infty}$H$^{+}_{2}$ ion has an internal structure which is completely 
different from usual atomic (i.e. one-center) three-body systems, e.g., from the two-electron H$^{-}$ ion and/or He atom. Indeed, for the two-center three-body systems the 
angular momentum of electron(s) ${\bf L}$ is not a constant of motion and cannot be used as a `good' quantum number. On the other hand, the operator $\Lambda = {\bf L}^2 + 
\sqrt{-2 E} R A_z$ which is responsible for separation of variables in spheroidal coordinates in the two-center ${}^{\infty}$H$^{+}_{2}$ ion, is not an integral of motion for 
atomic two-electron systems (see, e.g., \cite{Demk}, \cite{Fro1st} and references therein). Here the notation $A_z$ stands for the $z-$component of the Runge-Lenz operator 
(see, e.g., \cite{Lenz}), $R$ is the internuclear distance and $E$ is the total energy of the ${}^{\infty}$H$^{+}_{2}$ ion (a negative value). The same situation can be found 
with other quatum numbers which describe these two different groups of three-body systems. Briefly, we can say that the set of `good' quantum numbers used for the H$^{+}_{2}$ 
ion is fundamentally different from analogous sets of quantum numbers known for regular atomic three-body systems, e.g., for the two-electron H$^{-}$ ion, He atom and Ps$^{-}$ 
ion. 

Therefore, it is $a$ $priory$ clear that we cannot expect to reach a very high numerical accuracy in bound state calculations of the truly adiabatic system(s). Below we restrict 
ourselves to the analysis of the ground (bound) $1 s \sigma-$state in the ${}^{\infty}$H$^{+}_{2}$ ion. At this moment the best numerical result for the ground $1 s \sigma-$state 
energy in the ${}^{\infty}$H$^{+}_{2}$ ion is $\approx$ -0.60263119 $a.u.$ (see, e.g., \cite{JCP1} - \cite{JCP3} and \cite{Griv}). In \cite{Kut} the obtained total energy equals 
for this ion equals $\approx$ -0.602634214 $a.u.$ and this value agrees well with our recent results based on the use of cluster basis functions for two-center Coulomb systems.
These basis functions are very similar to our complex exxponents (see, Eqs.(\ref{equ75}) - (\ref{equ755}), but they also include an additional factors which essentially coincide 
with the `delta-distributions' written for the nucleus-nucleus $r_{21}$ (or $R$) coordinate. Results of our calculations will be published elswhere, but for our present purposes 
we can assume that the value -0.6026346(2) $a.u.$ is the very good approximation to the exact total energy. In this study we can try to obtain a relatively good approximation of 
this energy value. Note that for the model ${}^{\infty}$H$^{+}_{2}$ ion the Born-Oppenheimer approximation is not an approximation, but the exact approach which represents the 
actual structure of this system. Our universal variational expansions, Eqs.(\ref{equ75}) - (\ref{equ755}), have originally been developed for three-body systems with three 
comparable particle masses and for one-center atomic systems. For all these systems the exponential variational expansions, Eqs.(\ref{equ75}) - (\ref{equ755}), were found to be 
very effective. Therefore, we can expect that currently any direct calculation of the ground state energy in the ${}^{\infty}$H$^{+}_{2}$ ion with our variational expansion(s), 
Eqs.(\ref{equ75}) - (\ref{equ755}), will not be very accurate. The results of direct computations of the total energy of the ground state in the ${}^{M}$H$^{+}_{2}$ ion with very 
large values of the model `proton' mass $M = 1 \cdot 10^{28}$ $m_e$ (pure adiabatic system) can be found in Table IV (for different number of basis functions). As follows from 
this Table our results obtained for this model ion can be considered as accurate, but not highly accurate. However, the progress achieved just recently in calculations of the 
ground (bound) $1 s \sigma-$state of the ${}^{\infty}$H$^{+}_{2}$ ion is outstanding and very fast. Hopefully that after a few months of additional work we can present results of 
the first highly accurate variational computations of the model ${}^{\infty}$H$^{+}_{2}$ ion performed with our universal variational expansions, Eqs.(\ref{equ75}) - (\ref{equ755}). 

There is an alternative approach which can be used to reach our goal and produce an accurate numerical approximation to the ground $1 s \sigma$-state of the ${}^{\infty}$H$^{+}_{2}$ 
ion. In this method we approach the model ${}^{\infty}$H$^{+}_{2}$ ion by using a number of intermedium model $(a a e)^{+}$ ions (or ${}^{M}$H$^{+}_{2}$ ions) with the different $M$ 
values. In other words, we assume that the model ${}^{M}$H$^{+}_{2}$ ions contains the two very heavy nuclei with equal masses $M_1 = M_2 = M \gg m_e$. In our calculations of the 
ground $1 s \sigma-$state in the ${}^{\infty}$H$^{+}_{2}$ ions we consider a number of different (model) $(a a e)^{+}$ ions (or ${}^{M}$H$^{+}_{2}$ ions) with the following `proton' 
masses: $M = 6 \cdot 10^{3}$ $m_e$, $7 \cdot 10^{3}$ $m_e$, $\ldots$, $1 \cdot 10^{4}$ $m_e$ and $1.1 \cdot 10^{4}$ $m_e$. Highly accurate results of the total energy calculations 
(in atomic units) of these model ions in their ground $1 s \sigma-$states can be found in Table V. 

At the second step of this procedure, by using the results from Tables II and V for the symmetric $(a a e)^{+}$ ions, we can evaluate the total energy of the ground 
$1 s \sigma-$state in the truly adiabatic ${}^{\infty}$H$^{+}_{2}$ ion with the help of the following asymptotic formula
\begin{eqnarray}
  E(M) = E(\infty) + \sum^{K_a}_{k=2} C_k \Bigl(\frac{1}{M}\Bigr)^{\frac{k-1}{4}} =  E(\infty) + \sum^{K_a}_{k=2} C_k M^{\frac{1-k}{4}}
\end{eqnarray}
where $M$ is the mass of the model `proton' expressed in $m_e$, $K_a$ is the total number of terms used to approximate the numerical data obtained in a series of $K$ highly accurate 
calculations (with the different number of exponents $N$) of the total energies $E(M)$ (here we assume that $K_a \le K$). The derivation of this formula is also based on the 
Born-Oppenheimer approximation. By using our nine total energies from Tables II and V for the symmetric $(a a e)^{+}$ ions (i.e. $K$ = 9) we have found (for $K_a$ = 6) the following 
value of the total energy of the ground state in the ${}^{\infty}$H$^{+}_{2}$ ion $E(\infty) \approx$ -0.60263376559662592 $a.u.$ The deviation of this value from the known total 
energy of the ground state of the ${}^{\infty}$H$^{+}_{2}$ ion \cite{Griv} is $\approx 2.6 \cdot 10^{-6}$ $a.u.$ which is very good for our approximate procedure. 
  
\section{Conclusions}

We have explicitly constructed the universal variational expansion, Eq.(\ref{equ755}), which can be used to determine (to very high numerical accuracy) the bound states in various 
three-body systems, including Coulomb three-body systems with arbitrary particle masses and electric charges. Successful applications of our variational expansion now include: (1) 
atomic two-electron systems (He-atom(s) and He-like ions), (2) three-body with three comparable particle masses, and (3) molecularions with two heavy centers (or nuclei). 
Let us present here our numerical results (total energies) which have been obtained with the use of the variational expansion, Eq.(\ref{equ755}), for the three Coulomb three-body 
systems with unit charges, which are located in different parts of the `mass triangle' (its definition can be found in \cite{FroB92}). The total energy of the Ps$^{-}$ ion is $E$ = 
-0.26200507 02329801 07770400 315 $a.u.$, the total energy of the hydrogen negatively charged ion ${}^{\infty}$H$^{-}$ is $E$ = -0.527751 016544 377196 590573 $a.u.$ and the total 
energy of the hydrogen molecular ion ${}^{1}$H$^{+}_2$ (or H$^{+}_2$) is $E$ = -0.597139 063123 405074 8340(1) $a.u.$ (with the `old' particle masses, Table I) and $E$ = 
-0.597139 063175 573387 6052(1) $a.u.$ (with the `new' particle masses, Table II). The total energies of all ground states in the symmetric $(a a e)^{+}$ ions, i.e. for the 
$(p p e)^{+}, (d d e)^{+}$ and $(t t e)^{+}$ ions are now known to the same numerical accuracy (in contrast with our earlier studies \cite{Fro2002}). For the non-symmetric 
$(a b e)^{+}$ ions the progress achieved in this study is also amazing, since their total energies are now known to the accuracy which exceeds the overall accuracy of analogous 
results from \cite{Fro2002} by the factor of $2 \cdot 10^{4} - 5 \cdot 10^{5}$.

Formally, now we can say that the problem of highly accurate calculations of the bound states in arbitrary three-body systems is completely solved. Results obtained in this 
study for the adiabatic $(a a e)^{+}$ and $(a b e)^{+}$ two-center ions, where $(a, b) = (p, d, t)$, are unique and indicate clearly that our variational expansions, 
Eqs.(\ref{equ75}) - (\ref{equ755}), can effectively be used for highly accurate calculations of the bound states in these molecular ions and other similar systems. As follows 
from our results the same variational expansion(s) can be applied to perform highly accurate computations of the $(a a e)^{+}$ and $(a b e)^{+}$ ions with very large `proton' 
mass $M_p = \max( M_a, M_b) \approx 25,000 m_e$. This can be confirmed by our results from Table IV. Therefore, one can successfully apply our three-body variational expansion 
Eqs.(\ref{equ75}) - (\ref{equ755}) for the three-body adiabatic ions which contain the nuclei of helium, lithium and other heavy nuclei, e.g., for the one-electron 
$({}^{3}$He ${}^{1}$H$)^{2+}$ and $({}^{4}$He ${}^{2}$D$)^{2+}$ ions. Formally, the maximal `proton' mass $M_p \approx 25,000 m_e$ is sufficient to consider all actual 
three-body adiabatic systems with one bound electron. Furthermore, a few simple modifications of our current procedure can increase this value of $M_p$ up to $\approx$ 100,000 
- 250,000 $m_e$. However, we still need to reach a comparable numerical accuracy (e.g., $\pm 1 \cdot 10^{-17}$ $a.u.$) for the truly adiabatic ${}^{\infty}$H$^{+}_2$ ion and 
close two-center systems, e.g., in all model symmetric ions $(a a e)^{+}$, where $M_a \ge 250,000 m_e$. This must be achieved by using our universal variational expansions,
Eqs.(\ref{equ75}) - (\ref{equ755}), directly, i.e. without multiplying our basis functions by some additional `delta-distributions' for the nucleus-nucleus $r_{21}$ (or $R$) 
coordinate.    

Another important question which should be mentioned here is related to highly accurate computations of the bound state properties of the adiabatic three-body $(a a e)^{+}$ and 
$(a b e)^{+}$ ions which are of an increasing interest in a number of actual problems. Note that the expectation values of a large number of electron-nucleus and nucleus-nucleus 
properties of these adiabatic ions have been determined in \cite{Fro2002}. Our current expectation values for many bound state properties are now known to very high numerical
accuracy which is significantly better than accuracy reported in \cite{Fro2002}. In general, almost all bound state properties computed for the $(a a e)^{+}$ and $(a b e)^{+}$
ions are in very good agreement with the values computed in \cite{Fro2002}. Such an agreement is observed for all `regular' electron-nucleus and for most of the nucleus-nucleus 
expectation values. However, even with our improved wave functions we cannot produce a substantially better approximation for the `proton-proton' cusp values. Therefore, it is 
hard to evaluate the actual numerical accuracy of the computed `proton-proton' delta-functions $\langle \delta_{ab} \rangle$. Note that the `proton-proton' delta-functions 
$\langle \delta_{ab} \rangle$ determine the corresponding fusion rates in the $(a b e)^{+}$ molecular ions and other similar systems. Our current $\langle \delta_{ab} \rangle$ 
expectation values lead to some overestimation of these rates. Since 2002 the situation with these expectation values was improved significantly, but it is clear that additional 
research are needed in this area. Very likely, to prove the correctness of our expectation values of the nucleus-nucleus delta-functions we need to apply in our calculations the 
so-called `global identities' introduced by Drachman and Sucher in 1979 \cite{Dra1} and \cite{Dra2}. As it was shown earlier (see, e.g., \cite{Dra2} and \cite{Bha1} and 
references therein) these `global identities' allow one to obtain very accurate expectation values of all three delta-functions in an arbitrary three-body system. Right now, 
we are working to include such `global identities' in our calculations with the use of our exponential basis set with the complex non-linear parameters, Eq.(\ref{equ755}). 

Finally, we also want to emphasize that this our study is the first work where the `atomic' variational expansions in the relative and/or perimetric coordinates, 
Eq.(\ref{equ75}) - (\ref{equ755}), are applied for accurate numerical calculations of the ground bound state (or $1 s \sigma-$state) of the ${}^{\infty}$H$^{+}_{2}$ ion which is 
a pure adiabatic systems with two infinitely heavy Coulomb centers which do not move. As mentioned in the text above, originally, for the ${}^{\infty}$H$^{+}_{2}$ ion we could 
not expect to observe even a relatively good agreement between its `exact' energy \cite{JCP3} and the total energy obtained by our `universal' method. Nevertheless, right now 
our results are already in very good numerical agreement with the known `exact' energy. Furthermore, these results are rapidly improving, since, very likely, we have found an 
optimal strategy for optimization of our approach and variational wave function, Eq.(\ref{equ755}).  

\section{Acknowledgments}

Different parts of this project have been discussed with Yurii N. Demkov, David H. Bailey (Lawrence Berkeley National Laboratory), Tyson Whitehead (Sharcnet, UWO) and Styliani 
Constas (Dep. of Chemistry, UWO). Here I wish to thank all of them. All calculations for this project have been performed with the use of the extended arithmetic precision 
(software written and later modified by D.H. Bayley \cite{Bail1}, \cite{Bail2}). This work was supported in part by the NSF through a grant for the Institute for Theoretical 
Atomic, Molecular, and Optical Physics (ITAMP) at Harvard University and the Smithsonian Astrophysical Observatory. Also, I wish to thank James Babb (ITAMP), Gerry McKeon (UWO) 
and David M. Wardlaw (Memorial University) for stimulating discussions.

\newpage
\begin{table}[tbp]
   \caption{The total energies of the ground states (or $1 s \sigma-$states) of the $(p p e)^{+}, (p d e)^{+}, (p t e)^{+}, (d d e)^{+}, (d t e)^{+} $ and 
            $(t t e)^{+}$ molecular ions in atomic units. The notation $N$ is the total number of basis functions used.
            In these calculations the `old' particle masses have been used.}  
     \begin{center}
     \scalebox{0.85}{%
     \begin{tabular}{| c | c | c | c |}
      \hline\hline
 $N$  & $(p p e)^{+}$ (or H$_2^{+}$) & $(p d e)^{+}$ (or HD$^{+}$) & $(p t e)^{+}$ (or HT$^{+}$) \\  
     \hline
 2000 & -0.5971390631234050744620 & -0.5978979686450362041958 & -0.5981761346697656888411 \\
 2200 & -0.5971390631234050746908 & -0.5978979686450363954880 & -0.5981761346697660460870 \\
 2400 & -0.5971390631234050747785 & -0.5978979686450364682813 & -0.5981761346697661587977 \\
 2600 & -0.5971390631234050748083 & -0.5978979686450364940352 & -0.5981761346697662008105 \\
 2800 & -0.5971390631234050748237 & -0.5978979686450365059348 & -0.5981761346697662230799 \\
 3000 & -0.5971390631234050748283 & -0.5978979686450365117231 & -0.5981761346697662338929 \\
 3200 & -0.5971390631234050748312 & -0.5978979686450365149664 & -0.5981761346697662401499 \\ 
 3400 & -0.5971390631234050748321 & -0.5978979686450365164576 & -0.5981761346697662424531 \\
 3600 & -0.5971390631234050748326 & -0.5978979686450365178757 & -0.5981761346697662445535 \\
 3800 & -0.5971390631234050748331 & -0.5978979686450365184035 & -0.5981761346697662451284 \\
 4000 & -0.5971390631234050748334 & -0.5978979686450365187093 & -0.5981761346697662454937 \\
 4200 & -0.5971390631234050748336 & -0.5978979686450365189404 & -0.5981761346697662457764 \\
 4400 & -0.5971390631234050748337 & -0.5978979686450365190695 & -0.5981761346697662459529 \\
         \hline \hline
 $N$  & $(d d e)^{+}$  (or D$_2^{+}$) & $(t t e)^{+}$ (or T$_2^{+}$) & $(d t e)^{+}$ (or DT$^{+}$) \\ 
     \hline
 2000 & -0.5987887843306834639189 & -0.5995069101115414495185 & -0.5991306628550461721466 \\
 2200 & -0.5987887843306834643442 & -0.5995069101115414505292 & -0.5991306628550550695660 \\
 2400 & -0.5987887843306834644545 & -0.5995069101115414508826 & -0.5991306628550588287072 \\
 2600 & -0.5987887843306834645021 & -0.5995069101115414510357 & -0.5991306628550606468674 \\
 2800 & -0.5987887843306834645257 & -0.5995069101115414510956 & -0.5991306628550615293540 \\
 3000 & -0.5987887843306834645341 & -0.5995069101115414511183 & -0.5991306628550620750949 \\ 
 3200 & -0.5987887843306834645381 & -0.5995069101115414511322 & -0.5991306628550625897503 \\ 
 3400 & -0.5987887843306834645394 & -0.5995069101115414511364 & -0.5991306628550626110027 \\
 3600 & -0.5987887843306834645401 & -0.5995069101115414511392 & -0.5991306628550626248368 \\
 3800 & -0.5987887843306834645407 & -0.5995069101115414511408 & -0.5991306628550626275151 \\ 
 4000 & -0.5987887843306834645411 & -0.5995069101115414511416 & -0.5991306628550626286825 \\
 4200 & -0.5987887843306834645414 & -0.5995069101115414511423 & -0.5991306628550626296687 \\
 4400 & -0.5987887843306834645416 & -0.5995069101115414511425 & -0.5991306628550626301823 \\
         \hline \hline
  \end{tabular}}
  \end{center}
  \end{table}
\newpage
\begin{table}[tbp]
   \caption{The total energies of the ground states (or $1 s \sigma-$states) of the $(p p e)^{+}, (p d e)^{+}, (p t e)^{+}, (d d e)^{+}, (d t e)^{+}$ and 
            $(t t e)^{+}$ molecular ions in atomic units. the notation $N$ is the total number of basis functions used.
            In these calculations the `new' set of particle masses have been used.}  
     \begin{center}
     \scalebox{0.85}{%
     \begin{tabular}{| c | c | c | c |}
      \hline\hline
 $N$  & $(p p e)^{+}$ (or H$_2^{+}$) & $(p d e)^{+}$ (or HD$^{+}$) & $(p t e)^{+}$ (or HT$^{+}$) \\ 
     \hline   
 2000 & -0.5971390631755733873206 & -0.597897968692102925069 & -0.5981761346952079858529 \\         
 2200 & -0.5971390631755733875113 & -0.597897968692103129452 & -0.5981761346952083369879 \\
 2400 & -0.5971390631755733875621 & -0.597897968692103197261 & -0.5981761346952084606081 \\ 
 2600 & -0.5971390631755733875884 & -0.597897968692103225741 & -0.5981761346952085053123 \\ 
 2800 & -0.5971390631755733875979 & -0.597897968692103239394 & -0.5981761346952085313605 \\ 
 3000 & -0.5971390631755733876012 & -0.597879686921032443152 & -0.5981761346952085448930 \\ 
 3200 & -0.5971390631755733876028 & -0.597879686921032477262 & -0.5981761346952085540790 \\ 
 3400 & -0.5971390631755733876038 & -0.597879686921032486758 & -0.5991306628702067856575 \\ 
 3600 & -0.5971390631755733876042 & -0.597879686921032495322 & -0.5981761346952085585574 \\
 3800 & -0.5971390631755733876045 & -0.597879686921032500648 & -0.5981761346952085594235 \\
 4000 & -0.5971390631755733876047 & -0.597879686921032504200 & -0.5981761346952085598981 \\  
 4200 & -0.5971390631755733876049 & -0.597879686921032506605 & -0.5981761346952085601805 \\
 4400 & -0.5971390631755733876050 & -0.597879686921032507722 & -0.5981761346952085603146 \\
         \hline \hline
 $N$  & $(d d e)^{+}$ (or D$_2^{+}$) & $(t t e)^{+}$ (or T$_2^{+}$)  & $(d t e)^{+}$ (or DT$^{+}$) \\ 
     \hline
 2000 & -0.5987887843724103921958 & -0.5995069100946329213286 & -0.5991306628701917713791 \\
 2200 & -0.5987887843724103926244 & -0.5995069100946329220080 & -0.5991306628701998418175 \\
 2400 & -0.5987887843724103927496 & -0.5995069100946329222252 & -0.5991306628702031449100 \\
 2600 & -0.5987887843724103928172 & -0.5995069100946329223538 & -0.5991306628702048472878 \\ 
 2800 & -0.5987887843724103928451 & -0.5995069100946329223991 & -0.5991306628702057957716 \\
 3000 & -0.5987887843724103928553 & -0.5995069100946329224177 & -0.5991306628702062484756 \\
 3200 & -0.5987887843724103928614 & -0.5995069100946329224297 & -0.5991306628702067624721 \\
 3400 & -0.5987887843724103928630 & -0.5995069100946329224333 & -0.5991306628702067856575 \\
 3600 & -0.5987887843724103928638 & -0.5995069100946329224355 & -0.5991306628702067993622 \\
 3800 & -0.5987887843724103928644 & -0.5995069100946329224373 & -0.5991306628702068028691 \\
 4000 & -0.5987887843724103928649 & -0.5995069100946329224383 & -0.5991306628702068042175 \\
 4200 & -0.5987887843724103928653 & -0.5995069100946329224389 & -0.5991306628702068053301 \\
 4400 & -0.5987887843724103928655 & -0.5995069100946329224391 & -0.5991306628702068057470 \\
         \hline \hline
  \end{tabular}}
  \end{center}
  \end{table}
%
\begin{table}[tbp]
   \caption{The total energies of the ground states (or $1 s \sigma-$states) of the ground states of the $(p p e)^{+}, (p d e)^{+}, (p t e)^{+}, (d d e)^{+}, 
            (d t e)^{+}, (t t e)^{+}$ and ${}^{\infty}$H$^{+}_{2}$ ions (in atomic units) determined with the short-term cluster
            wave functions. The notation $N$ is the total number of basis functions used.}  
     \begin{center}
     \begin{tabular}{| c | c | c | c | c | c |}
      \hline\hline
  ion & $N$ & `new' masses & `old' masses & N & ${}^{\infty}$H$^{+}_{2}$ \\ 
                        \hline \hline
  $(p p e)^{+}$ & 400 & -0.597139031755653 & -0.597139009308909 & 200 & -0.602565528163790 \\ 

  $(p d e)^{+}$ & 400 & -0.597897965889470 & -0.597897965878425 & 400 & -0.602595877443168 \\

  $(p t e)^{+}$ & 400 & -0.598176129563061 & -0.598176130448024 & 600 & -0.602596747201453 \\

  $(d t e)^{+}$ & 400 & -0.599130655125588 & -0.599130650680771 & 700 & -0.602597398839590 \\

  $(d d e)^{+}$ & 400 & -0.598788756393092 & -0.598788757197200 & 800 & -0.602597496538961 \\

  $(t t e)^{+}$ & 400 & -0.599506882560035 & -0.599506885299115 & --- & ---------------- \\
      \hline\hline
  \end{tabular}
  \end{center}
  \end{table}
%
\begin{table}[tbp]
   \caption{The total energies $E$ of the ground $1 s \sigma-$state in the ${}^{\infty}$H$^{+}_{2}$ ion (in atomic units). 
            The notation $N$ is the total number of basis functions used. The `proton' mass used in these calculations equals 
            $1 \cdot 10^{28}$ $m_e$.}  
     \begin{center}
     \begin{tabular}{| c | c | c | c |}
      \hline\hline
 $N$  & $E$ & $N$ & $E$ \\ 
     \hline
  400 & -0.602595877443168 & 2000 & -0.602629173786511 \\
 1000 & -0.602618755343140 & 2400 & -0.602630245875453 \\
 1600 & -0.602627834161557 & 2800 & -0.602631579156735 \\
         \hline \hline
  \end{tabular}
  \end{center}
  \end{table}
%
\begin{table}[tbp]
   \caption{The total energies of the ground states (or $1 s \sigma-$states) in the $(a a e)^{+}$ ions (in atomic units). 
            The notation $M$ stands for the mass of the model `proton', while symbol $N$ is the total 
            number of basis functions used.}  
     \begin{center}
     \scalebox{0.85}{%
     \begin{tabular}{| c | c | c |}
      \hline\hline
 $N$  & $M = 6 \cdot 10^{3} m_e$  & $M = 7 \cdot 10^{3} m_e$ \\ 
     \hline
 1800 & -0.59964364285569943581 & -0.59986971956219381454 \\
 2000 & -0.59964364285569943800 & -0.59986971956219381913 \\
 2200 & -0.59964364285569943900 & -0.59986971956219382111 \\
 2400 & -0.59964364285569943932 & -0.59986971956219382180 \\
 2600 & -0.59964364285569943946 & -0.59986971956219382212 \\
 2800 & -0.59964364285569943952 & -0.59986971956219382227 \\
 3000 & -0.59964364285569943955 & -0.59986971956219382235 \\ 
 3200 & -0.59964364285569943956 & -0.59986971956219382241 \\
 4000 & -0.59964364285569943958 & -0.59986971956219382246 \\ 
 4200 & -0.59964364285569943958 & -0.59986971956219382246 \\ 
         \hline \hline
 $N$  & $M = 8 \cdot 10^{3} m_e$  & $M = 9 \cdot 10^{3} m_e$ \\ 
     \hline
 1800 & -0.60005146723865326314 & -0.60020168153488045022 \\
 2000 & -0.60005146723865327738 & -0.60020168153488050145 \\
 2200 & -0.60005146723865328293 & -0.60020168153488052029 \\
 2400 & -0.60005146723865328495 & -0.60020168153488052786 \\
 2600 & -0.60005146723865328618 & -0.60020168153488053303 \\
 2800 & -0.60005146723865328675 & -0.60020168153488053545 \\
 3000 & -0.60005146723865328712 & -0.60020168153488053717 \\ 
 3200 & -0.60005146723865328760 & -0.60020168153488053810 \\
 4000 & -0.60005146723865328763 & -0.60020168153488053974 \\
         \hline \hline
 $N$  & $M = 1 \cdot 10^{4} m_e$  & $M = 1.1 \cdot 10^{4} m_e$ \\ 
     \hline
 1800 & -0.60032852464778396510 & -0.60043747991110986735 \\
 2000 & -0.60032852464778413957 & -0.60043747991111039833 \\
 2200 & -0.60032852464778420302 & -0.60043747991111059276 \\
 2400 & -0.60032852464778423167 & -0.60043747991111069074 \\
 2600 & -0.60032852464778425131 & -0.60043747991111075566 \\
 2800 & -0.60032852464778426068 & -0.60043747991111078763 \\
 3000 & -0.60032852464778426777 & -0.60043747991111081276 \\
 3200 & -0.60032852464778427860 & -0.60043747991111085287 \\
 4000 & -0.60032852464778427883 & -0.60043747991111085355 \\
      \hline\hline
  \end{tabular}}
  \end{center}
  \end{table}
\end{document}